\documentclass[prd,preprint,aps,superscriptaddress]{revtex4}
\usepackage[utf8]{inputenc}
\usepackage{amsmath}
\usepackage{amssymb}
\usepackage{amsfonts}
\usepackage{textcomp}
\usepackage{float}
\usepackage{ulem} 
\usepackage[dvips]{graphicx}
\usepackage{slashed}
\usepackage{tikz}

\newcommand\brabar{\scalebox{.3}{(}\raisebox{-1.7pt}[0pt][0pt]{$-$}\scalebox{.3}{)}}

\preprint{FERMILAB-PUB-19-185-T}

\begin{document}
\title{A Study of the PDF uncertainty on the LHC W-boson mass measurement}
\author{Moh'd Hussein}
\email{m.hussein@ju.edu.jo}
\affiliation{Department of Physics, The University of Jordan, Amman 11942, Jordan}
\author{Joshua Isaacson}
\email{isaacson@fnal.gov}
\affiliation{Theoretical Physics Department, Fermilab, Batavia, IL 60510, USA}
\author{Joey Huston}
\email{huston@pa.msu.edu}
\affiliation{Department of Physics and Astronomy, Michigan State University, East Lansing, MI 48824, USA}
\renewcommand{\abstractname}{}
\begin{abstract}
  The determination of the $W$-boson mass through an analysis of the decay charged-lepton transverse momentum distribution
  has a sizable uncertainty due to the imperfect knowledge of the relevant parton distribution functions (PDFs).
  In this paper, a quantitative assessment of the $W$-boson mass uncertainty at the LHC resulting from the PDF uncertainty is examined.  
  We use the \texttt{CT14 NNLO} PDFs with a  NNLL + NNLO calculation (\texttt{ResBos}) to simulate the $W$-boson production and
  decay kinematics.  The uncertainty of the $W$-boson mass determination is then determined as a function of the boson and lepton kinematics. 
 For $W^{+}$ production using $P_{T}^{W} < 15$ GeV and
 $35 < P_{T}^{l}$ (GeV) $< 45$, PDF uncertainties (at the 68\% CL) of $^{+16.0}_{-17.5}$ MeV,  $^{+13.9}_{-14.8}$ MeV, and  $^{+12.2}_{-19.2}$ MeV, are determined at 7 TeV, 8 TeV and 13 TeV respectively.
The uncertainties of $W^{-}$ for the same cuts are found to be  $^{+15.9}_{-15.6}$ MeV, $^{+15.0}_{-12.7}$ MeV and $^{+14.8}_{-15.3}$ MeV,
    at 7 TeV, 8 TeV and 13 TeV respectively.     
\end{abstract}

\maketitle

\section{Introduction}

\subsection{Importance of the precision $W$-boson mass determination} 
 
The $W$-boson mass ($M_{W}$) is a fundamental parameter of the Standard Model (SM). $M_{W}$ can be both directly measured, and computed indirectly from electroweak precision constraints. There is an interplay between $M_{W}$,
the top quark mass ($M_{t}$), and the Higgs mass ($M_{H}$) when calculating electroweak precision observables~\cite{Baak:2012kk}. Currently, there is some tension between the best fit value from the electroweak precision data and the direct measurement.
Electroweak precision tests are most sensitive to $M_{W}$. 
If a statistically significant discrepancy between the indirect measurement of $M_{W}$ and the direct measurement is found, this would be strong evidence for new physics beyond the SM such as the Minimal Supersymmetric Standard Model~\cite{Heinemeyer:2006px}.  
Therefore, more precise determinations of $M_{W}$ have been an important goal for collider physics programs at both the Tevatron and the LHC, both for testing the consistency of the SM and for exploring the possibility of new physics beyond the SM\@. 

\subsection{Current limits on $M_{W}$ and on PDF uncertainty}

Currently, the best limits for the $W$-boson mass come from the measurements at the Tevatron, and from a recent result from ATLAS\@.
The $W$-boson mass has been measured at CDF (80387$\pm$19 MeV) and D0 ($80367 \pm 26$ GeV), with a  
 combined Tevatron average of 80387$\pm$16 MeV~\cite{Aaltonen:2013vwa, D0:2013jba, Aaltonen:2013iut}. A more recent 
 measurement for $M_{W}$ at ATLAS has yielded a value of $80370 \pm 19$ MeV~\cite{Aaboud:2017svj}. CMS has not reported
 a result to date.
 One of the major systematic uncertainties for the direct measurement arises from an imperfect knowledge of the parton distribution 
 functions (PDFs) relevant for $W$-boson production.
 Using the $M_{H}$ measurement~\cite{Aad:2015zhl}, the combined value for $M_{t}$~\cite{ATLAS:2014wva}, and the SM precisely measured
 parameters of the fine-structure constant $\alpha$, the Fermi constant $G_{\mu}$, and the $Z$-boson mass ($M_{Z}$),
 results in a predicted (indirect) mass for the $W$-boson of 80362$\pm$8 MeV~\cite{deBlas:2016ojx} and 80358$\pm$8 MeV~\cite{Baak:2014ora}.
 This indirect uncertainty limit provides a goal for the desired experimental precision for the W-boson mass. 

\subsection{PDFs and their uncertainties}

Parton distribution functions cannot at present be calculated from first principles, but must be determined by data. This determination requires the use of data from a wide variety of processes and experiments, both for the determination of the central PDFs and of their uncertainties. 
 There are several collaborations dedicated to this endeavor, among them:
 \texttt{CTEQ-TEA}~\cite{Dulat:2015mca}, \texttt{MMHT}~\cite{Harland-Lang:2014zoa}, and
 \texttt{NNPDF}~\cite{Ball:2014uwa}. PDFs have been produced at LO, NLO and (more recently) NNLO, in the strong coupling constant $\alpha_s$.
 The highest precision predictions at the LHC 
 require the use of NNLO PDFs. Previously, NNLO PDFs from the CTEQ collaboration (\texttt{CT10}~\cite{Gao:2013xoa}), have been used
 in a determination of the W-mass uncertainty~\cite{Bozzi:2015hha}.
 In this paper, we update those predictions using the NNLO PDFs of \texttt{CT14}~\cite{Dulat:2015mca}. 
 The \texttt{CT14} PDFs include data sets  from the LHC at $\sqrt{s}$ = 7 TeV for the first time, 
 as well as updated data from the Tevatron and 
 from  HERA\@. The most important aspect  of the LHC data sets for this analysis is their ability to  impose constraints on 
 the light quark and anti-quark PDFs at parton x-values appropriate for W-boson production at the LHC\@.  

PDF uncertainties are calculated for \texttt{CT14} through the use of Hessian eigenvectors~\cite{Pumplin:2001ct}.  \texttt{CT14 NNLO} has 56 
Hessian error PDFs, corresponding to 28 eigenvectors, which result from the 28 free PDF parameters in the \texttt{CT14} fit.
In comparison, \texttt{CT10 NNLO} had 50 error sets, resulting from using 25 free parameters. The increased number of parameters
is a result of a more flexible parametrization used in the \texttt{CT14} global PDF fit. 
  The purpose of the present analysis is to provide a 
 quantitative assessment of the $M_{W}$ uncertainty resulting from PDF uncertainties using the  \texttt{CT14 NNLO}  
 PDFs.   

\section{Methods for determining $M_{W}$}

The production of W/Z bosons is one of the most well-studied examples of hard scattering processes
at hadron colliders~\cite{Alioli:2016fum}.
After production, the W boson decays into jets, or into a lepton-neutrino pair. Decays into the former channel are difficult to resolve due to 
large contributions from QCD dijet background processes. On the other hand, the $W \rightarrow e\nu_{e}$ and 
$W \rightarrow \mu\nu_{\mu}$ channels 
allow for precise measurements of prompt, energetic, and isolated
electrons and muons. The decay of $W \rightarrow \tau\nu_{\tau}$ is also not included for precision measurements due to the complex nature of the $\tau$ decay.

 The only observables directly measured by the detectors are the momenta of 
 the leptons ($P^{l}$) and of the hadrons ($P^{\text{hadrons}}$) produced in association with the $W$-boson; the latter is referred to as the hadronic recoil. In addition, the hadronic recoil's transverse momentum defines  
 the negative transverse momentum of the $W$-boson itself ($P_{T}^{W}$). 
Since the 
 neutrinos escape the detector without interaction, $M_{W}$ cannot be reconstructed on an event-by-event basis. 
 However, the sum of the transverse momentum of all particles in the event should sum to zero in the absence of any 
 particles evading detection and detector resolution effects. Therefore, the neutrino's transverse momentum
 ($P_{T}^{\nu}$) can be inferred indirectly from the transverse energy missing from 
 the event, $\slashed{E}_{T}\equiv - (P_{T}^{l} + P_{T}^{\text{hadrons}}$). 
 The transverse mass of the $W$-boson ($M_{T}^{W}$), introduced in~\cite{Barger:1983wf,Smith:1983aa}, is defined as: 
 \begin{equation}
     M_{T}^{W}=\sqrt{2 P_{T}^{l} \slashed{E}_{T}(1-\cos \Delta\phi_{l\nu})},
  \end{equation}   
  where ($\Delta\phi_{l\nu}$) is the azimuthal angle between the lepton and the neutrino (or the missing transverse energy).

It is potentially possible to measure the $W$-boson mass using any one of the three kinematic variables,
 $M_{T}^{W}$,
 $P_{T}^{l}$ and
 $P_{T}^{\nu}$.
 In practice, the experimental resolution for $P_{T}^{\nu}$ does not allow for a competitive
measurement of the W-boson mass (but it can still be useful as a cross-check). The other two variables can, and have been,  used though at both the Tevatron and the LHC.

 The extraction of $M_{W}$ is obtained through the use of templates generated from a highly parametrized Monte
 Carlo (MC) simulation. At leading order (LO), the $W$-boson is produced with zero transverse momentum,
 and thus the charged lepton and neutrino are always back-to-back. Therefore, in the LO calculation, in the limit of zero width, and with a perfect detector,
 $M_{T}^{W}$ and $P_{T}^{l}$ would have extremely sharp Jacobian peaks exactly at $M_{W}$ 
 and $M_{W}/2$ respectively. QCD radiation, and the impact of detector resolution, results in a shift in the location of the Jacobian peak and a broadening the distributions. 

\subsection{Differences between Tevatron and LHC}\label{errors}

Compared to the  Tevatron, the LHC experiments benefit from larger signal and calibration samples. For luminosities in
 the multi-$fb^{-1}$ range, the data samples are larger by an order of magnitude compared to the corresponding samples used at the Tevatron, and 
thus the statistical errors are significantly smaller. Moreover, and given the precisely measured value of
$M_{Z} (91187.6 \pm 2.1\ \text{MeV})$~\cite{Abbaneo:2000nr}
 and the clean leptonic
 final state, the $Z \rightarrow l^{+}l^{-}$ processes are able to be used to model the detector's response 
 to $W \rightarrow l\nu$ and to validate the analysis synopsis~\cite{Krasny:2007cy}.

 Uncertainties in the PDFs are the dominant source of error for the extraction of $M_{W}$ at
 the Tevatron (a complete list of these uncertainties can be found in Tables XIV and VI in
 Ref.~\cite{D0:2013jba,Aaltonen:2013vwa} respectively).  
 The PDF uncertainties for $W$-boson production were expected to be larger at the LHC~\cite{Krasny:2010zz},
 due to the smaller parton x-values (where the uncertainties are larger) being sampled, and due to larger contributions from sea quarks from a $pp$ collider vs.\ a $p\bar{p}$ collider (see Table 3 of Ref.~\cite{Aaboud:2017svj}).
 
 Whereas $W$-boson production at the Tevatron is charge symmetric, the $W^{+}$-boson production rate at the LHC exceeds 
 that of $W^{-}$-boson by about 40\%. Moreover, the second generation quarks contribute 
 only  approximately 5\% of the overall $W$-boson production rate at the Tevatron, while at the LHC ($\sqrt{s}$ = 7 TeV) this rate is approximately 
 25\% of the overall $W$-boson production rate~\cite{Aaboud:2017svj}. This fraction continues to increase as the center-of-mass energy increases. The uncertainty on the strange and charm quarks is larger than those on the light quarks(anti-quarks), and thus result in a proportionally larger contribution to the $W$ boson mass determination. 

Compared to $P_{T}^{l}$, the 
$M_{T}^{W}$ measurement at the LHC is affected by larger experimental systematic uncertainties,
due, for example, to high pile-up energy deposited in the detector %~\cite{pile-up}
from the additional proton-proton interactions in each bunch crossing. 
This results in a degradation of  the resolution of the measurement of 
$P_{T}^{\text{hadrons}}$ that scales roughly as the square root of the total hadronic energy in the event~\cite{ATL-PHYS-PUB-2014-015}.
The limiting factor to balance, at present, between the experimental uncertainty ($M_{T}^{W}$) 
and the theoretical uncertainty ($P_{T}^{l}$) depends on the ability to develop pile-up 
mitigation techniques~\cite{Aad:2015ina}. Although both techniques have been used at the LHC~\cite{Aaboud:2017svj},
the greater discriminatory power lies with the use of the $P_{T}^{l}$ distribution, as shown in
Table 10 and in Fig. 23 of Ref.~\cite{Aaboud:2017svj}. Therefore, this paper will focus on determining the PDF uncertainties for the $P_{T}^{l}$ distribution. 

\subsection{Resummed QCD}

At NLO, the $W$-boson can recoil against one parton, and at NNLO, against two partons, thus acquiring a non-zero transverse momentum.
For $P_{T}^{W}$ much less than $M_{W}$, soft gluon radiation has a large impact on the
$P_{T}^{W}$, and resummation calculations are necessary to provide a good description of the distribution.
At very low $P_{T}$ values, non-perturbative effects also become important, and must also be taken into account in
the resummation calculations.  
In practice, most $W$-boson events used in the determination of $M_{W}$~\cite{Aaboud:2017svj}
are produced with low $P_{T}^{W}$ values (smaller than 30 GeV).
This means that, in principle, a resummed generator with non-perturbative effects such as \texttt{ResBos} is preferable to fixed-order calculations~\cite{Quackenbush:2015yra}.

Fixed-order QCD predictions work well for $W$ boson production at intermediate and large $P_{T}$.  However, the $W/Z$-bosons are predominantly
 produced at low $P_{T}$, where the fixed-order cross section behaves as 
\begin{equation}
    \frac{d \sigma}{d P^{2}_{T}} \sim \frac{1}{P^{2}_{T}}\sum_{n=1}^\infty {}_{k}c_{n} \alpha_{s}^{n}\left(Q\right)\sum_{k=0}^{2n-1} \ln^{k} \frac{P^{2}_{T}}{Q^{2}},
\end{equation}  
in the limit of $P_T \rightarrow 0$. It can be seen that the above equation
has an unphysical divergence when $\ln^{k}(P^{2}_{T}/Q^{2}) \rightarrow 0$. This results in large logarithms ($\alpha_s^n\ln^{k}(P^{2}_T/Q^2) > 1$), 
making perturbative calculations unreliable~\cite{Campbell:2017hsr}.
This unphysical divergence is addressed through the
resummation procedure, in which the logarithmic terms are resummed to all orders in the $\alpha_{s}$ expansion. 
The general formalism for transverse momentum resummation was first introduced by Collins, Soper, and Sterman (CSS)~\cite{Collins:1984kg}. 
\begin{equation}
    \frac{d\sigma}{dQ^2dydq_t^2}= \sigma_0 \left( \int \frac{d^2b}{{(2\pi)}^2}e^{i q_t \cdot b} \widetilde{W}(b) + Y\right),
\end{equation}
where
\begin{equation}
\widetilde {W}(b)=e^{-{\cal S}_{pert}(Q^2,b_*)-S_{NP}(Q,b)}\sum_{i,j} C_{qi}\otimes f_{i/A}(x_1,\mu=C_0/b_*) C_{\bar qj}\otimes f_{j/B}(x_2,\mu=C_0/b_*).
\end{equation}
Here $\sigma_0$ is the leading order cross section, $Y$ is the regular piece in the limit $q_t \rightarrow 0$, 
the $C$'s contain the hard collinear virtual corrections, $S_{pert}$ is the perturbative Sudakov factor, 
$S_{NP}$ is the non-perturbative Sudakov factor, $x_{1,2}=\frac{Q}{\sqrt{s}}e^{\pm y}$ represent the momentum 
fractions carried by the incoming partons in the given process, and $f_{i/A}$,$f_{j/B}$ are the PDFs. The non-perturbative Sudakov factor
is introduced to handle the Landau pole in QCD in the limit that $b \rightarrow \infty$, or $\mu\rightarrow 0$.
Therefore, we adopt the $b_*$ formalism introduced in~\cite{Collins:1984kg}, and defined as:
 \begin{equation}
     b_*=\frac{b}{\sqrt{1+b^2/b_{\text{max}}^2}}  \ ,\quad b_{\text{max}}<1/\Lambda_{QCD}\ \label{bstar} 
\end{equation}
The resummed and fixed-order calculations then have to be matched at intermediate $P_{T}$, where the fixed-order calculation does not contain large logarithms, in order to obtain QCD predictions
for the entire range of $P_{T}$. Kinematic restrictions on the decay products of the $W$-boson can then be applied to mimic the cuts applied to the data. 

\subsection{\texttt{ResBos} versus parton shower}

\texttt{ResBos}, the analytic resummation program proposed and discussed in detail in Refs.~\cite{Ladinsky:1993zn,Balazs:1997xd,Landry:2002ix}, 
is used to calculate at next-to-next-to-leading-log and next-to-next-to-leading order (NNLL) + (NNLO)
the W boson cross section for the process $pp\rightarrow W^{\pm}+X\rightarrow l^{\pm}\overset{\brabar}{\nu}+X$,
using a renormalization and factorization scale of $\mu_{R/F}=M_{W}$.
\texttt{ResBos} uses a NNLO/NLO k-factor to obtain 
the NNLO correction for the $Y$-piece of the resummed cross-section. This k-factor is calculated as a function of $Q, q_t$, and $y$.
The non-perturbative Sudakov form factor, which describes the $P_{T}^{W}$ at
low $P_{T}$, is fit using DY data, and parameterized by the BLNY form~\cite{Landry:2002ix}. 

Parton shower Monte Carlo (MC) programs such as \texttt{POWHEG}~\cite{Alioli:2008gx} can also be used to simulate $W$-boson production at the LHC\@.
Parton showers resum the leading tower of logarithms while resummation programs can include higher order logarithms,
providing a more accurate description of the $P_{T}^{W}$ distribution. 
The $W$-boson decay distributions are highly sensitive to the order of resummation in the calculation,
and these small differences can potentially lead to an uncertainty in $M_{W}$ greater than the desired goal of 10 MeV.
Therefore, it is important to use a calculation which includes as many higher logs and higher order corrections as possible
to obtain the highest accuracy currently obtainable. 

The MC event generators output are fully exclusive, i.e.\ they provide information on all final-state particles,  
but approximate the resummation effects through the parton shower. On the other hand, resummation programs are inclusive, that is they provide information only about the $W$-boson and its decay products, integrating out all additional QCD radiation.
However, the information provided by resummation codes is sufficient for a precise analysis of the $W$-boson mass.
Additionally, corrections to the width of $W$-boson and spin correlations between the initial and final state 
particles are included in \texttt{ResBos} when applicable.

\section{Technique for PDFs error uncertainty determination}\label{chi2}

\subsection{PDF uncertainty estimation}

 In order to quantify the PDF uncertainty on $M_{W}$, we need to fit the detailed shapes of the $P_{T}^{l}$ distributions
 using the log-likelihood ($\chi^{2}$) analysis.
 The steps we have followed are outlined below, following Ref.~\cite{Bozzi:2015hha}: 

\begin{itemize}
\item[1 -] While keeping $M_{W}$ fixed at a given central value ($M_{W,0}$=80358 MeV), 
we generate the lepton distribution $P_{T}^{l}$ for the 56 Hessian error PDFs.
Here we choose the mass at 80358 MeV based on the indirect fit obtained in Ref.~\cite{Baak:2014ora} (since we are only interested 
in the shift of the mass due to the PDF, and not what the central prediction is, this is an acceptable
choice to make).
\item[2 -] We now leave the PDF fixed to the central value (CT14), and vary $M_{W}$ in 
the generator to obtain the lepton transverse momentum distributions $(P_{T}^{l})$
for each of the different masses considered. 
Here we consider values between 80308 MeV and 80408 MeV, creating templates in steps of 1 MeV. 
\item[3 -] For each template generated in step 1, corresponding to the different error PDFs for a fixed W boson mass (termed $P^{\text{PDF}}_{i}$), 
    we loop through all the different templates generated in step 2, corresponding to the different $W$-boson masses using the central CT14 PDF (termed $P^{\text{Mass}}_{j}$),
and compute the corresponding $\chi^{2}_{ij}$;
\begin{equation}
    \chi^{2}_{ij}=\sum_{k=1}^{N_{\text{bins}}}\frac{{(P^{\text{PDF}}_{i}-P^{\text{Mass}}_{j})}^{2}}{{(\sigma^{\text{PDF}}_{i})}^{2}+{(\sigma^{\text{Mass}}_{j})}^{2}},
\end{equation}
where the summation is over all of the lepton transverse momentum bins and $\sigma$ is the statistical uncertainty for the given bin. 
\item[4 -] For a fixed PDF $j$, the value of $i$ that minimizes the $\chi^2_{ij}$ distribution corresponds to
the mass that would be predicted by the given PDF error set. In other words, the mass
predicted by PDF $j$ would be mass $i$ ($M_{W,i}$), if the corresponding $\chi^2_{ij}$ was the smallest. 
\end{itemize}

The PDF Hessian uncertainty on $M_{W}$ then is calculated using the master equation proposed in Ref~\cite{Nadolsky:2001yg,Lai:2010vv} for asymmetric uncertainties
as follows:
\begin{equation}
\begin{split}
    \Delta M_{W}^{+} =\sqrt{\sum_{i=1}^{28} {[\max (\{M_{W,i^{+}}-M_{W,0}\},\{M_{W,i^{-}}-M_{W,0}\}, 0)]}^{2}}
\\
\Delta M_{W}^{-} =\sqrt{\sum_{i=1}^{28} {[\max (\{M_{W,0}-M_{W,i^{+}}\},\{M_{W,0}-M_{W,i^{-}}\}, 0)]}^{2}}
\end{split}
\end{equation}
where $M_{W,i}^{\pm}$ represents the best fit value for the $M_{W}$ corresponding to the PDF set
$i$ induced by a change of $\pm$ 1 standard deviations of each independent parameter describing the PDF set. 

Normalizing the $P_{T}^{l}$ templates to use the shape of the distribution, instead of the overall rate,
in the fit region can substantially reduce the size of the
PDF uncertainties without losing the sensitivity to the value of $M_{W}$~\cite{ATL-PHYS-PUB-2014-015}. Thus, in this study we use normalized distributions. 

\subsection{Fit parameters and kinematic cuts}

\texttt{ResBos} is used to calculate the $W^{\pm}$ 
boson kinematics at NNLL+NNLO for the processes $pp\rightarrow W^{\pm}+X\rightarrow l^{\pm}\overset{\brabar}{\nu}+X$, based on the 
\texttt{CT14 NNLO} sets, at $\sqrt[]{s}$ = 7, 8 and 13 TeV. A charm pole mass of 1.3 GeV is used, as in the \texttt{CT10 NNLO} PDFs. 
The PDFs for up, down, strange (anti) quarks and the gluon are parametrized at an initial scale of 1.295 GeV. The central 
PDF sets are obtained using a central value of $\alpha_{s}(M_{Z})$ of 0.118, as recommended by the 
\texttt{PDF4LHC} group~\cite{Butterworth:2015oua}. The \texttt{CT14 NNLO} PDFs uncertainties are provided as 90\% confidence level (CL) intervals,
 and then are scaled by a multiplicative factor of (1/1.642) to provide 68\% CL intervals. 
This scaling is appropriate if the $\chi^2$ distribution is suitably quadratic. 

 The $W$-boson signal in the data is extracted by selecting events with one central isolated, high 
 $P_{T}^{l}$ lepton, large missing energy, and low hadronic recoil. In the \texttt{ResBos}-generated events, the  cuts described below are implemented, mimicking the cuts used in the experimental analyses. The missing transverse energy ($\slashed{E}_{T}$)
 is required to be greater than 20 GeV and the absolute value of the lepton pseudo-rapidity ($|\eta|$)
 is restricted to the region less than 2.5. 
For comparative purposes, different ranges
 for $P_{T}^{l}$ (in GeV) are used: ($35 < P_{T}^{l} < 45,\ 30 < P_{T}^{l} < 50$,
 and $20 < P_{T}^{l} < 60$). All of these ranges have equally spaced bins of 0.5 GeV.
 Additional boson transverse momentum cuts
 ($P_{T}^{W} < 15$ GeV, $P_{T}^{W} < 30$ GeV,
 $P_{T}^{W} < 60$ GeV and $P_{T}^{W} < 300$ GeV) are applied as well for comparison.
We generate approximately 100M events for each template to minimize the effects of statistical fluctuations.
 
\section{Results}

\begin{figure}[ht]
\begin{center}
\includegraphics[width=0.45\textwidth, angle=0]{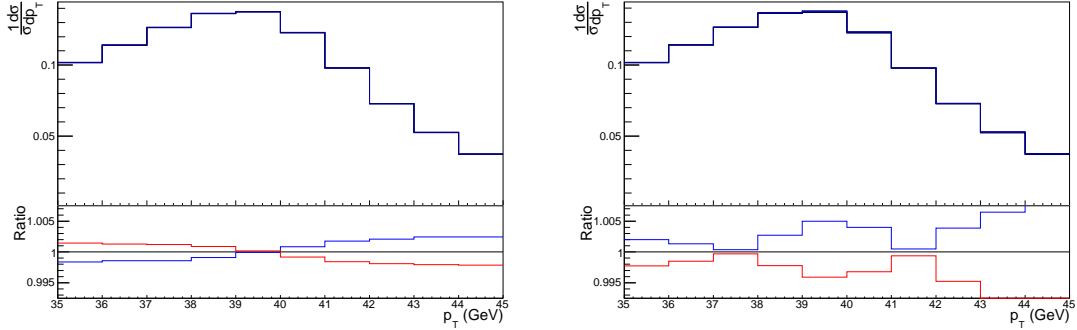}
\includegraphics[width=0.45\textwidth, angle=0]{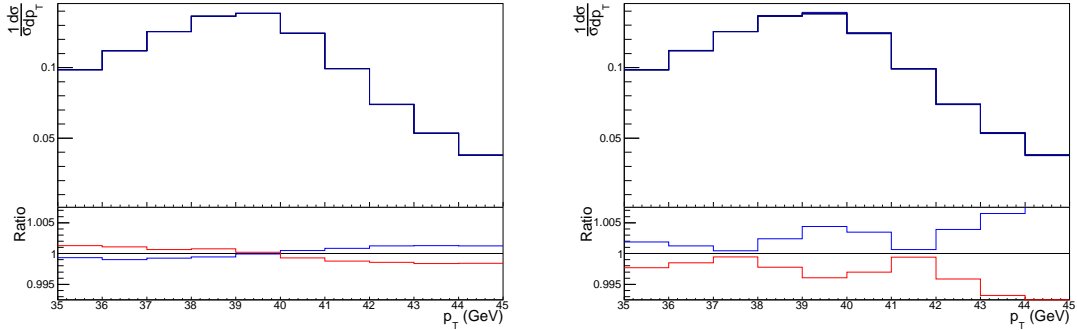}
\caption{$P_{T}^{l}$ (in GeV) for the different $P^{\text{Mass}}_{j}$ (left)
  and $P^{\text{PDF}}_{i}$ (right) PDF errors for $W^{+}$ at $\sqrt[]{s}$ = 7 TeV.}\label{fig:7plus}
\end{center}
\end{figure}

\begin{figure}[ht]
\begin{center}
\includegraphics[width=0.45\textwidth, angle=0]{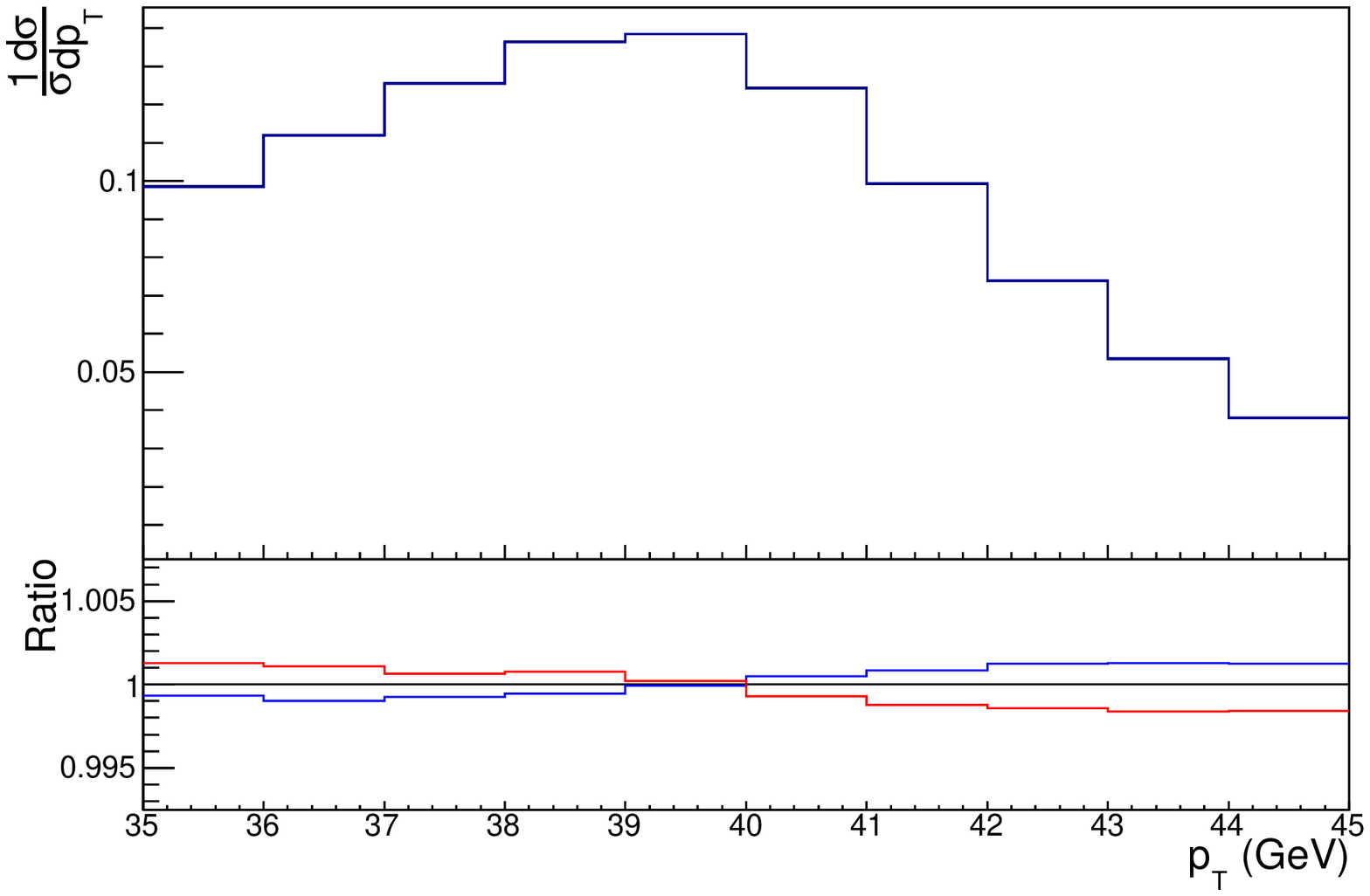}
\includegraphics[width=0.45\textwidth, angle=0]{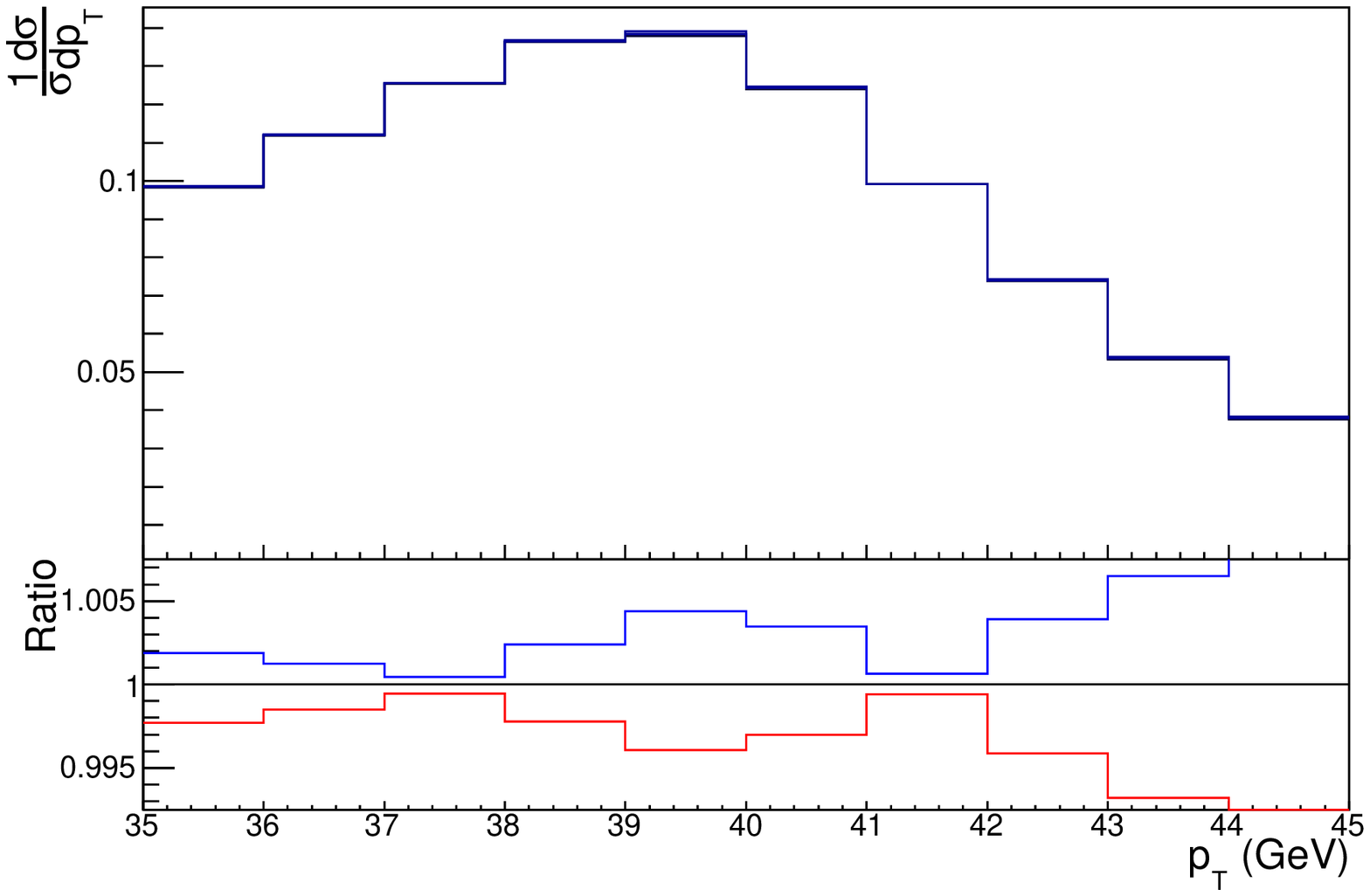}
\caption{$P_{T}^{l}$ (in GeV) for the different $P^{\text{Mass}}_{j}$ (left)
  and $P^{\text{PDF}}_{i}$ (right) PDF errors for $W^{-}$ at $\sqrt[]{s}$ = 7 TeV.}\label{fig:7minus}
\end{center}
\end{figure}

Figure~\ref{fig:7plus} shows sample lepton transverse momentum distributions for the $W$-boson mass variations (left) and for the error PDF variations (right) for $W^{+}$ at at $\sqrt[]{s}$ = 7 TeV. Figure~\ref{fig:7minus} shows the same lepton transverse momentum distributions for $W^{-}$.  The variations, too small to easily observe in the top distributions, are more easily seen in the ratio plots underneath. 

For the ratio plots for the W boson mass variations (left side of Figs.~\ref{fig:7plus},\ref{fig:7minus}), the red (blue) curve corresponds to the maximum (minimum) allowed mass from the $\chi^2$ fit. The crossing of the two curves occurs due to the fact that the result is normalized to unity. Similarly, the ratio plots for the PDF variations (right side of Figs.~\ref{fig:7plus},\ref{fig:7minus}), the upper (lower)  curve corresponds to the total positive (negative) PDF uncertainty obtained from the master equation
for the transverse momentum of the lepton. Here the curves do not cross due to the fact that the upper curve always corresponds to the positive direction uncertainty, while the lower corresponds to the negative direction uncertainty. For the variation of the 
error PDFs, the positive and negative ratios are approximately mirror images of each other, indicating that the uncertainty on the $W$ mass from the PDFs is reasonably symmetriic. 

\begin{figure}[ht]
\begin{center}
\includegraphics[width=0.80\textwidth, angle=0]{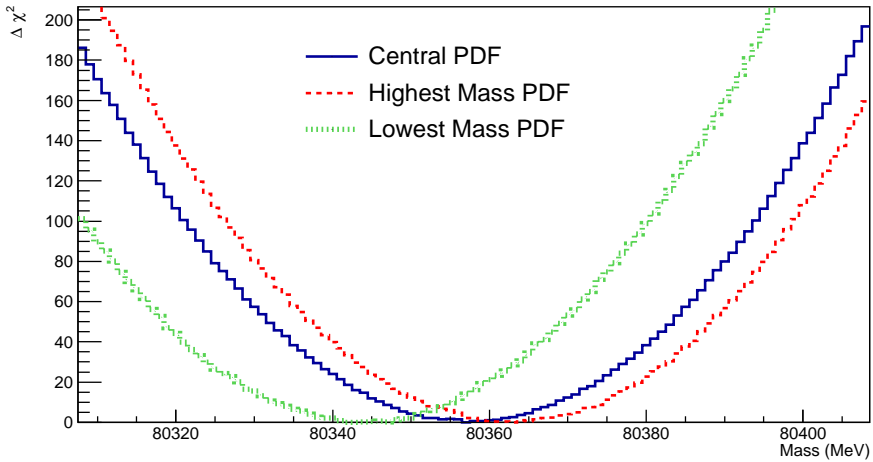}
\caption{The $\chi^2_{ij}$ profiles as a function of $M_{W}$ for $W^{+}$ at $\sqrt[]{s}$ = 7 TeV.}
\label{fig:Chi2}
\end{center}
\end{figure}

As an example, the $\Delta\chi^2_{ij}$ profiles as a function of the $W^{+}$-boson mass at $\sqrt[]{s}$ = 7 TeV are shown in Figure~\ref{fig:Chi2}. The $\Delta\chi^2_{ij}$ profiles are reproduced for the central PDF, for the PDF that produces the highest mass $W$-boson, and the PDF that produces the lowest mass $W$-boson. Note that the $\Delta\chi^2_{ij}$ distributions are parabolic, indicating that 
the 90\% errors can be scaled down to 68\% by applying a scaling factor.
Similar curves are obtained at center-of-mass energies of 8 and 13 TeV. 

The Hessian PDF uncertainties, as a function of various kinematic cuts, for both $W^{\pm}$ at $\sqrt[]{s}$ = 7 TeV, are listed in Table~\ref{table:7}.  Those for $\sqrt[]{s}$ = 8 TeV are listed in Table~\ref{table:8}, and those for $\sqrt[]{s}$ = 13 TeV are shown in Table~\ref{table:13}.

A number of generalizations can be made from this information. 
As expected, restricting the $W$-boson transverse momentum, $P_{T}^{W}$, to be \textless 15 GeV results in the smallest PDF uncertainties,
  as does restricting the $P_{T}^{l}$ range to a narrow band  ($35  < P_{T}^{l}$
  (GeV) $ < 45$) about the Jacobian peak. The PDF uncertainty (and the $+/-$ asymmetry for the uncertainty) tends to grow as the $P_{T}^{W}$ cut increases.  The PDF uncertainties are similar 
for $W^{+}$ and $W^{-}$ production. 

 A previous analysis~\cite{Bozzi:2015hha}, using the techniques outlined here, estimated the PDF uncertainty for the $W$ mass
 using a selection of PDF sets including \texttt{CT10 NNLO}. The analysis framework was based on the \texttt{POWHEG} 
 MC interfaced  with \texttt{PYTHIA} PS~\cite{Sjostrand:2014zea}. This analysis obtained a global uncertainty that ranged between $\pm 18$ and $\pm 24$ MeV, 
 depending on the final state, collider energy and the PDF set. For \texttt{CT10 NNLO}, Ref.~\cite{Bozzi:2015hha}
 reported that at $\sqrt[]{s}$ = 13 TeV and with $P_{T}^{W} < 15$ GeV and 
 $29 < P_{T}^{l}$ (GeV) $< 49$, the errors are $^{+20}_{-17}$ MeV and $^{+17}_{-9}$ MeV for
 $W^{+}$ and $W^{-}$ respectively. 
 The results in this paper for a similar lepton transverse momentum range (30 \textless P$_{\mbox{\tiny T}}^{\mbox{\tiny l}}$ (GeV) \textless 50)
 are $^{+17.8}_{-18.3}$ MeV and $^{+15.7}_{-17.1}$ MeV respectively.
 At $\sqrt[]{s}$ = 8 TeV, Ref.~\cite{Bozzi:2015hha} reported errors of $^{+17}_{-18}$ MeV for $W^{+}$ and $^{+16}_{-11}$ MeV for
 $W^{-}$, similar to what we have observed.

\begin{table}[ht]  
\centering
\begin{tabular}{|c|c|c|}
\hline
$35 < P_{T}^{l}\ \text{(GeV)}\ < 45$ &  $W^{+}$ Uncertainty (MeV) &  $W^{-}$ Uncertainty (MeV)\\
\hline \hline
$P_{T}^{W} < 15$ GeV & + 16.0 - 17.5 & + 15.9  - 15.6\\ 
\hline
$P_{T}^{W} < 30$ GeV & + 18.1 - 22.4 & + 22.4  - 18.1\\
\hline
$P_{T}^{W} < 60$ GeV & + 14.2 - 24.7 & + 21.1  - 20.5\\
\hline
$P_{T}^{W} < 300$ GeV &+ 12.5 - 32.3 & + 28.5  - 17.2\\
\hline \hline
$30 < P_{T}^{l}\ \text{(GeV)}\ < 50$ &  $W^{+}$ Uncertainty (MeV) &  $W^{-}$ Uncertainty (MeV)\\
\hline \hline
$P_{T}^{W} < 15$ GeV & + 17.4 - 18.7 & + 15.0  - 17.7 \\
\hline
$P_{T}^{W} < 30$ GeV & + 20.1 - 27.6 & + 23.0  - 24.6\\
\hline
$P_{T}^{W} < 60$ GeV & + 21.1 - 27.8 & + 19.4  - 29.7\\
\hline
$P_{T}^{W} < 300$ GeV & + 17.8 - 30.0 & + 19.2  - 29.3\\
\hline \hline
$20 < P_{T}^{l}\ \text{(GeV)}\ < 60$ &  $W^{+}$ Uncertainty (MeV) &  $W^{-}$ Uncertainty (MeV)\\
\hline \hline
$P_{T}^{W} < 15$ GeV & + 20.0 - 19.3 & + 20.7  - 14.1 \\
\hline
$P_{T}^{W} < 30$ GeV & + 23.1 - 31.3 & + 31.9  - 21.5\\
\hline
$P_{T}^{W} < 60$ GeV & + 19.9 - 39.4 & + 14.1  - 42.8\\
\hline
$P_{T}^{W} < 300$ GeV & + 17.5 - 43.8 & + 10.0  - 51.3\\
\hline
\end{tabular}
\caption{The impact of different cuts on the PDFs uncertainty of $M_{W}$ at $\sqrt[]{s}$ = 7 TeV.}
\label{table:7}
\end{table}

\begin{table}[ht]  
\centering
\begin{tabular}{|c|c|c|}
\hline
$35 < P_{T}^{l} \text{(GeV)} < 45$ & $W^{+}$ Uncertainty (MeV) & $W^{-}$ Uncertainty (MeV)\\
\hline \hline
$P_{T}^{W} < 15$ GeV & + 13.9 - 14.8 & + 15.0  - 12.7\\ 
\hline
$P_{T}^{W} < 30$ GeV & + 21.7 - 16.3 & + 18.9  - 17.2\\
\hline
$P_{T}^{W} < 60$ GeV & + 21.5 - 21.7 & + 22.4  - 19.7\\
\hline
$P_{T}^{W} < 300$ GeV &+ 20.0 - 25.2 & + 26.3  - 15.9\\
\hline \hline
$30 < P_{T}^{l} \text{(GeV)} < 50$ & $W^{+}$ Uncertainty (MeV) & $W^{-}$ Uncertainty (MeV)\\
\hline \hline
$P_{T}^{W} < 15$ GeV & + 16.1 - 16.1 & + 17.6  - 15.4 \\
\hline
$P_{T}^{W} < 30$ GeV & + 23.5 - 20.2 & + 23.0  - 21.2\\
\hline
$P_{T}^{W} < 60$ GeV & + 24.5 - 21.8 & + 29.1  - 19.7\\
\hline
$P_{T}^{W} < 300$ GeV & + 21.8 - 23.5 & + 28.3  - 20.0\\
\hline \hline
$20 < P_{T}^{l} \text{(GeV)} < 60$ & $W^{+}$ Uncertainty (MeV) & $W^{-}$ Uncertainty (MeV)\\
\hline \hline
$P_{T}^{W} < 15$ GeV & + 13.6 - 21.9 & + 17.9  - 16.6 \\
\hline
$P_{T}^{W} < 30$ GeV & + 20.9 - 32.0 & + 28.7  - 21.9\\
\hline
$P_{T}^{W} < 60$ GeV & + 25.1 - 34.5 & + 32.4  - 27.9\\
\hline
$P_{T}^{W} < 300$ GeV & + 34.2 - 26.8 & + 34.7  - 27.8\\
\hline
\end{tabular}
\caption{The impact of different cuts on the PDFs uncertainty of $M_{W}$ at $\sqrt[]{s}$ = 8 TeV.}
\label{table:8}
\end{table}

\begin{table}[ht]  
\centering
\begin{tabular}{|c|c|c|}
\hline
$35 < P_{T}^{l}\ \text{(GeV)}\ < 45$ & $W^{+}$ Uncertainty (MeV) & $W^{-}$ Uncertainty (MeV)\\
\hline \hline
$P_{T}^{W} < 15$ GeV & + 12.2 - 19.2 & + 14.8  - 15.3\\ 
\hline
$P_{T}^{W} < 30$ GeV & + 16.3 - 25.9 & + 14.7  - 24.8\\
\hline
$P_{T}^{W} < 60$ GeV & + 19.5 - 27.1 & + 20.7  - 24.3\\
\hline
$P_{T}^{W} < 300$ GeV &+ 21.8 - 27.6 & + 16.4  - 27.3\\
\hline \hline
$30 < P_{T}^{l}\ \text{(GeV)} < 50$ & $W^{+}$ Uncertainty (MeV) & $W^{-}$ Uncertainty (MeV)\\
\hline \hline
$P_{T}^{W} < 15$ GeV & + 17.8 - 18.3 & + 15.7  - 17.1 \\
\hline
$P_{T}^{W} < 30$ GeV & + 24.2 - 24.6 & + 20.2  - 25.4\\
\hline
$P_{T}^{W} < 60$ GeV & + 24.0 - 25.8 & + 23.8  - 26.7\\
\hline
$P_{T}^{W} < 300$ GeV & + 20.4 - 29.3 & + 21.0  - 27.6 \\
\hline \hline
$20 < P_{T}^{l} \text{(GeV)} < 60$ & $W^{+}$ Uncertainty (MeV) & $W^{-}$ Uncertainty (MeV)\\
\hline \hline
$P_{T}^{W} < 15$ GeV & + 18.2 - 18.2 & + 16.9  - 17.2 \\
\hline
$P_{T}^{W} < 30$ GeV & + 27.6 - 28.2 & + 23.6  - 27.1\\
\hline
$P_{T}^{W} < 60$ GeV & + 27.5 - 33.7 & + 30.4  - 30.6\\
\hline
$P_{T}^{W} < 300$ GeV & + 26.4 - 33.5 & + 24.6  - 30.3\\
\hline
\end{tabular}
\caption{The impact of different cuts on the PDFs uncertainty of $M_{W}$ at $\sqrt[]{s}$ = 13 TeV.}
\label{table:13}
\end{table}

\section{Summary}

The data samples for W-boson production at the LHC are large, for all center-of-mass energies. 
However, one of the limiting factors for determination of $M_{W}$ is not statistics,
but rather our  imprecise knowledge of PDFs. In this study, we investigated this dominant source of uncertainty.

To improve  the PDF uncertainty on the $M_{W}$ determination, a better knowledge of the relevant PDFs is needed. Such improvement may be possible as new generations of PDFs (such as CT18) include more LHC data, such as precision measurements of the $W$ and $Z$ boson cross sections.

\section*{Acknowledgment}
We would like to than M. Boonekamp and A. Vicini for useful discussions, and for C.$-$P. Yuan for suggesting the project.
This work was partly supported by Fermi Research Alliance, LLC, under Contract 
  No.\ DE--AC02--07CH11359 with the U.S.\ Department of Energy, Office of 
  Science, Office of High Energy Physics. 
\bibliography{WMass}

\end{document}